# Mechanisms behind large Gilbert damping anisotropies

I. P. Miranda[1], A. B. Klautau[2,*], A. Bergman[3], D. Thonig[3,4], H. M. Petrilli[1], and O. Eriksson[3,4]

[1]*Universidade de São Paulo, Instituto de Física,*
*Rua do Matão, 1371, 05508-090, São Paulo, SP, Brazil*
[2]*Faculdade de Física, Universidade Federal do Pará, Belém, PA, Brazil*
[3]*Department of Physics and Astronomy, Uppsala University, Box 516, SE-75120 Uppsala, Sweden and*
[4]*School of Science and Technology, Örebro University, Fakultetsgatan 1, SE-701 82 Örebro, Sweden*

(Dated: November 22, 2021)

A method with which to calculate the Gilbert damping parameter from a real-space electronic structure method is reported here. The anisotropy of the Gilbert damping with respect to the magnetic moment direction and local chemical environment is calculated for bulk and surfaces of $Fe_{50}Co_{50}$ alloys from first principles electronic structure in a real space formulation. The size of the damping anisotropy for $Fe_{50}Co_{50}$ alloys is demonstrated to be significant. Depending on details of the simulations, it reaches a maximum-minimum damping ratio as high as 200%. Several microscopic origins of the strongly enhanced Gilbert damping anisotropy have been examined, where in particular interface/surface effects stand out, as do local distortions of the crystal structure. Although theory does not reproduce the experimentally reported high ratio of 400% [Phys. Rev. Lett. 122, 117203 (2019)], it nevertheless identifies microscopic mechanisms that can lead to huge damping anisotropies.

*Introduction:* Magnetic damping has a critical importance in determining the lifetime, diffusion, transport and stability of domain walls, magnetic vortices, skyrmions, and any nano-scale complex magnetic configurations [1]. Given its high scientific interest, a possibility to obtain this quantity by means of first-principles theory [2] opens new perspectives of finding and optimizing materials for spintronic and magnonic devices [3–8]. Among the more promising ferromagnets to be used in spintronics devices, cobalt-iron alloys demonstrate high potentials due to the combination of ultralow damping with metallic conductivity [4, 9].

Recently, Li *et al.* [10] reported an observed, giant anisotropy of the Gilbert damping ($\alpha$) in epitaxial $Fe_{50}Co_{50}$ thin films (with thickness $10-20$ nm) reaching maximum-minimum damping ratio values as high as 400%. The authors of Ref. [10] claimed that the observed effect is likely due to changes in the spin-orbit coupling (SOC) influence for different crystalline directions caused by short-range orderings that lead to local structural distortions. This behaviour differs distinctly from, for example, pure bcc Fe [11]. In order to quantitatively predict the Gilbert damping, Kambersky's breathing Fermi surface (BFS) [12] and torque-correlation (TC) [13] models are frequently used. These methods have been explored for elements and alloys, in bulk form or at surfaces, mostly via reciprocal-space *ab-initio* approaches, in a collinear or (more recently) in a noncollinear configuration [14]. However, considering heterogeneous materials, such as alloys with short-range order, and the possibility to investigate element specific, non-local contributions to the damping parameter, there are, to the best of our knowledge, no reports in the literature that rely on a real space method.

In this Letter, we report on an implementation of *ab initio* damping calculations in a real-space linear muffin-tin orbital method, within the atomic sphere approximation (RS-LMTO-ASA) [15, 16], with the local spin density approximation (LSDA) [17] for the exchange-correlation energy. The implementation is based on the BFS and TC models, and the method (Supplemental Material - SM, for details) is applied to investigate the reported, huge damping anisotropy of $Fe_{50}Co_{50}(100)/MgO$ films [10]. A main result here is the identification of a microscopic origin of the enhanced Gilbert damping anisotropy of $Fe_{50}Co_{50}(100)$ films, and the intrinsic relationships to the local geometry of the alloy. Most significantly, we demonstrate that a surface produces extremely large damping anisotropies that can be orders of magnitude larger than that of the bulk. We call the attention to the fact that this is the first time, as far as we know, that damping values are theoretically obtained in such a local way.

*Results:* We calculated: *i)* ordered $Fe_{50}Co_{50}$ in the $B2$ structure (hereafter refereed to as $B2$-FeCo) *ii)* random $Fe_{50}Co_{50}$ alloys in bcc or bct structures, where the virtual crystal approximation (VCA) was applied; *iii)* $Fe_{50}Co_{50}$ alloys simulated as embedded clusters in a VCA matrix (host). In all cases VCA was simulated with an electronic concentration corresponding to $Fe_{50}Co_{50}$. The *ii)* and *iii)* alloys were considered as in bulk as well as in the (001) surface, with bcc and bct structures (hereafter correspondingly refereed as VCA $Fe_{50}Co_{50}$ bcc, VCA $Fe_{50}Co_{50}$ bct, VCA $Fe_{50}Co_{50}(001)$ bcc and VCA $Fe_{50}Co_{50}(001)$ bct). The effect of local tetragonal distortions was considered with a local $\frac{c}{a} = 1.09$ ratio (SM for details). All data for cluster based results, were obtained from an average of several different configurations. The total damping for a given site $i$ in real-space ($\alpha_t$, Eqs. S6 and S7 from SM) can be decomposed in non-local, $\alpha_{ij}$ ($i \neq j$), and local (onsite), $\alpha_{onsite}$ (or $\alpha_{ii}$, $i = j$) contributions, each of them described by the tensor elements

$$\alpha_{ij}^{\nu\mu} = \frac{g}{m_i \pi} \int \eta(\epsilon) \text{Tr}\left(\hat{T}_i^\nu \text{Im} \hat{G}_{ij} \hat{T}_j^\mu \text{Im} \hat{G}_{ji}\right) d\epsilon, \quad (1)$$

where $m_i$ is the total magnetic moment localized in the reference atomic site $i$, $\mu, \nu = \{x, y, z\}$, $\hat{T}$ is the torque operator, and $\eta(\epsilon) = \frac{\partial f(\epsilon)}{\partial \epsilon}$ the derivative of the Fermi distribution. The scalar $\alpha_{ij}$ parameter is defined in the collinear regime as $\alpha_{ij} = \frac{1}{2}(\alpha_{ij}^{xx} + \alpha_{ij}^{yy})$.

To validate our methodology, the here obtained total damping for several systems (such as bcc Fe, fcc Ni, hcp and fcc Co and $B2$-FeCo) were compared with established values available in the literature (Table S1, SM), where an overall good agreement can be seen.

Fig. 1 shows the non-local contributions to the damping for bcc Fe and $B2$-FeCo. Although the onsite contributions are around one order of magnitude larger than the non-local, there are many $\alpha_{ij}$ to be added and total net values can become comparable. Bcc Fe and $B2$-FeCo have very different non-local damping contributions. Element resolved $\alpha_{ij}$, reveal that the summed Fe-Fe interactions dominate over Co-Co, for distances until $2a$ in $B2$-FeCo. We observe that $\alpha_{ij}$ is quite extended in space for both bcc Fe and $B2$-FeCo. The different contributions to the non-local damping, from atoms at equal distance arises from the reduced number of operations in the crystal point group due to the inclusion of SOC in combination with time-reversal symmetry breaking. The $B2$-FeCo arises from replacing every second Fe atom in the bcc structure by a Co atom. It is interesting that this replacement (i.e. the presence of Co in the environment) significantly changes the non-local contributions for Fe-Fe pairs, what can more clearly be seen from the *Inset* in Fig. 1, where the non-local damping summed over atoms at the same relative distance for Fe-Fe pairs in bcc Fe and $B2$-FeCo are shown; the non-local damping of Fe-Fe pairs are distinctly different for short ranges, while long ranged (further than $\sim 2.25$ Å) contributions are smaller and more isotropic.

The damping anisotropy, i.e. the damping change, when the magnetization is changed from the easy axis to a new direction is[1]

$$\Delta \alpha_t = \left( \frac{\alpha_t^{[110]}}{\alpha_t^{[010]}} - 1 \right) \times 100\%, \quad (2)$$

where $\alpha_t^{[110]}$ and $\alpha_t^{[010]}$ are the total damping obtained for magnetization directions along [110] and [010], respectively. Analogous definition also applies for $\Delta \alpha_{onsite}$.

---

[1] We note that this definition is different to the maximum-minimum damping ratio, defined as $\frac{\alpha_t^{[110]}}{\alpha_t^{[010]}} \times 100\%$, from Ref. [10].

We investigated this anisotropy in surfaces and in bulk systems with (and without) tetragonal structural distortions. Our calculations for VCA $Fe_{50}Co_{50}$ bcc show a damping increase of $\sim 13\%$, when changing the magnetization direction from [010] to [110] (Table S2 in the SM). The smallest damping is found for the easy magnetization axis, [010], which holds the largest orbital moment ($m_{orb}$) [18]. For VCA $Fe_{50}Co_{50}$ bcc we obtained a small variation of $\sim 2\%$ for the onsite contribution ($\alpha_{onsite}^{[010]} = 8.94 \times 10^{-4}$ and $\alpha_{onsite}^{[110]} = 8.76 \times 10^{-4}$), what implies that the anisotropy comes mostly from the non-local contributions, particularly from the next-nearest neighbours. For comparison, $\Delta \alpha_t \sim 3\%$ (with $\Delta \alpha_{onsite} \sim 0.4\%$) in the case of bcc Fe, what corroborates the reported [11] small bcc Fe anisotropy at room temperature, and with the bulk damping anisotropy rates [19].

We also inspected the chemical inhomogeneity influence on the anisotropy, considering the $B2$-FeCo alloy, where the weighted average damping (Eq. S7 of SM) was used instead. The $B2$-FeCo bcc ($\sim 7\%$) and VCA $Fe_{50}Co_{50}$ bcc ($\sim 13\%$) anisotropies are of similar magnitudes. Both $B2$ structure and VCA calculations lead to damping anisotropies which are significantly lower than what was observed in the experiments, and it seems likely that the presence of disorder in composition and/or structural properties of the Fe/Co alloy would be important to produce large anisotropy effects on the damping.

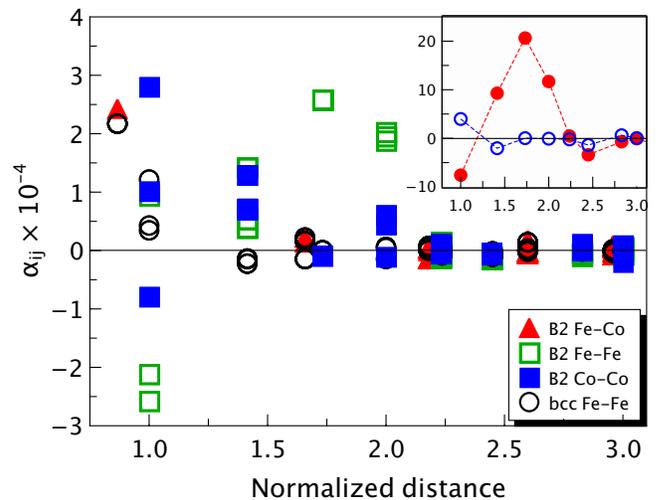

Figure 1. (Color online) Non-local damping contributions, $\alpha_{ij}$, in (Fe-centered) bulk $B2$-FeCo and bcc Fe, as a function of the normalized distance in lattice constant units $a$. *Inset*: Non-local contributions from only Fe-Fe pairs summed, for each distance, in bcc Fe bulk (empty blue dots) and in the $B2$-FeCo (full red dots). The onsite damping for Fe (Co) in $B2$-FeCo is $\alpha_{onsite}^{Fe} = 1.1 \times 10^{-3}$ ($\alpha_{onsite}^{Co} = 0.8 \times 10^{-3}$) and for bcc Fe it is $\alpha_{onsite}^{Fe} = 1.6 \times 10^{-3}$. The magnetization direction is $z$ ([001]). Lines are guides for the eyes.

We analyzed the role of local distortions by considering a hypothetical case of a large, 15% ($\frac{c}{a} = 1.15$), distortion on the $z$-axis of ordered $B2$-FeCo. We found the largest damping anisotropy ($\sim 24\%$) when comparing the results with magnetization in the [001] ($\alpha_t^{[001]} = 10.21 \times 10^{-3}$) and in the [010] ($\alpha_t^{[010]} = 7.76 \times 10^{-3}$) directions. This confirms that, indeed, bct-like distortions act in favour of the $\Delta\alpha_t$ enhancement (and therefore, of the maximum-minimum damping ratio), but the theoretical data are not large enough to explain the giant value reported experimentally [10].

Nevertheless, in the case of an alloy, the local lattice distortions suggested in Ref. [10] are most to likely occur in an heterogeneous way [20], with different distortions for different local environments. To inspect this type of influence on the theoretical results, we investigated (Table S3, SM) clusters containing different atomic configurations embedded in a VCA $Fe_{50}Co_{50}$ matrix (with Fe bulk lattice parameter); distortions were also considered such that, locally in the clusters, $\frac{c}{a} = 1.15$ (Table S4, in the SM). Moreover, in both cases, two types of clusters have to be considered: Co-centered and Fe-centered. The $\alpha_t$ was then computed as the sum of the local and non-local contributions for clusters with a specific central (Fe or Co) atom, and the average of Fe- and Co-centered clusters was taken. Fe-centered clusters have shown larger anisotropies, on average $\sim 33\%$ for the undistorted ($\sim 74\%$ for the distorted) compared with $\sim 8\%$ for the undistorted Co-centered clusters ($\sim 36\%$ for the distorted). Although these results demonstrate the importance of both, local distortions as well as non-local contributions to the damping anisotropy, they are not still able to reproduce the huge observed [10] maximum-minimum damping ratio.

We further proceed our search for ingredients that could lead to a huge $\Delta\alpha_t$ by inspecting interface effects, which are present in thin films, grain boundaries, stacking faults and materials in general. Such interfaces may influence observed properties, and in order to examine if they are relevant also for the reported alloys of Ref. [10], we considered these effects explicitly in the calculations. As a model interface, we considered a surface, what is, possibly, the most extreme case. Hence, we performed a set of $\alpha_t$ calculations for the $Fe_{50}Co_{50}$(001), first on the VCA level. Analogous to the respective bulk systems, we found that the onsite contributions to the damping anisotropy are distinct, but they are not the main cause ($\Delta\alpha_{onsite} \sim 18\%$). However, the lack of inversion symmetry in this case gives a surprisingly large enhancement of $\Delta\alpha_t$, thus having its major contribution coming from the non-local damping terms, in particular from the next-nearest neighbours. Interestingly, negative non-local contributions appear when $\alpha_t$ is calculated in the [010] direction. These diminish the total damping (the onsite contribution being always positive) and gives rise to a larger anisotropy, as can be seen by comparison of the results shown in Table I and Table S5 (in the Supplemental Material). In this case, the total anisotropy was found to be more than $\sim 100\%$ (corresponding to a maximum-minimum damping ratio larger than 200%).

A compilation of the most relevant theoretical results obtained here is shown in Fig. 2, together with the experimental data and the local density of states (LDOS) at $E_F$ for each magnetization direction of a typical atom in the outermost layer (data shown in yellow). As shown in Fig. 2, the angular variation of $\alpha_t$ has a fourfold ($C_{4v}$) symmetry, with the smallest Gilbert damping occurring at 90° from the reference axis ([100], $\theta_H = 0°$), for both surface and bulk calculations. This pattern, also found experimentally in [10], matches the in-plane bcc crystallographic symmetry and coincides with other manifestations of SOC, such as the anisotropic magnetoresistance [10, 21]. Following the simplified Kambersky's formula [13, 22], in which (see SM) $\alpha \propto n(E_F)$ and, therefore, $\Delta\alpha \propto \Delta n(E_F)$, we can ascribe part of the large anisotropy of the FeCo alloys to the enhanced LDOS differences at the Fermi level, evidenced by the close correlation between $\Delta n(E_F)$ and $\Delta\alpha_t$ demonstrated in Fig. 2. Thus, as a manifestation of interfacial SOC (the so-called proximity effect [23]), the existence of $\Delta\alpha_t$ can be understood in terms of Rashba-like SOC, which has been shown to play an important role on damping anisotropy [24, 25]. Analogous to the bulk case, the higher $m_{orb}$ occurs where the system presents the smallest $\alpha_t$, and the orbital moment anisotropy matches the $\Delta\alpha_t$ fourfold symmetry with a 90° rotation phase (see Fig. S3, SM). Note that a lower damping anisotropy than $Co_{50}Fe_{50}$(001) is found for a pure Fe(001) bcc surface, where it is $\sim 49\%$ (Table S2, SM), in accordance with Refs. [7, 26], with a dominant contribution from the onsite damping values (conductivity-like character on the reciprocal-space [19, 27]).

The VCA surface calculations on real-space allows to investigate the layer-by-layer contributions (intra-layer damping calculation), as shown in Table I. We find that the major contribution to the damping surface anisotropy comes from the outermost layer, mainly from the difference in the minority $3d$ states around $E_F$. The deeper layers exhibit an almost oscillatory $\Delta\alpha_t$ behavior, similar to the oscillation mentioned in Ref. [28] and to the Friedel oscillations obtained for magnetic moments. The damping contributions from deeper layers are much less influenced by the inversion symmetry breaking (at the surface), as expected, and eventually approaches the typical bulk limit. Therefore, changes in the electronic structure considered not only the LDOS of the outermost layer but a summation of the LDOS of all layers (including the deeper ones), which produces an almost vanishing difference between $\theta_H = 0°$ and $\theta_H = 45°$ (also approaching the bulk limit). The damping anisotropy arising as a surface effect agrees with what was observed in the case of Fe [7] and CoFeB [29] on GaAs(001), where the damping

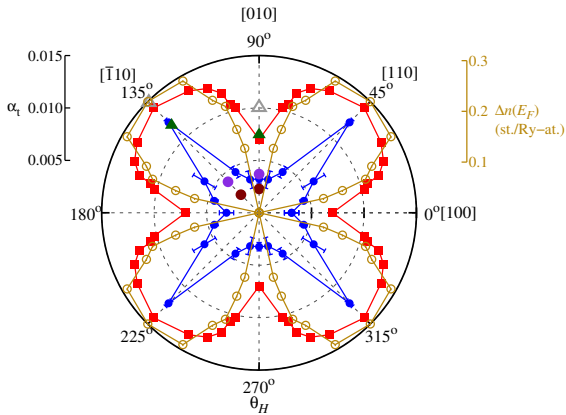

Figure 2. (Color online) Total damping and LDOS difference at $E_F$, $\Delta n(E_F)$, as a function of $\theta_H$, the angle between the magnetization direction and the [100]-axis. Squares: (red full) VCA $Fe_{50}Co_{50}$(001) bcc. Triagles: (green full) average over 32 clusters (16 Fe-centered and 16 Co-centered), with bcc structure at the surface layers (SM) embedded in a VCA medium; (gray open) similar calculations, but with a local lattice distortion. Circles: (yellow open) $\Delta n(E_F)$ between $\theta_H = 0°$ and the current angle for a typical atom in the outermost layer of VCA $Fe_{50}Co_{50}$(001) bcc; (blue full) experimental data [10] for a 10-nm $Fe_{50}Co_{50}$/Pt thin film; (purple full) average bulk VCA $Fe_{50}Co_{50}$ bcc; and (brown full) the $B2$-FeCo bulk. Lines are guides for the eyes.

anisotropy diminishes as the film thickness increases.

Table I. Total intra-layer damping ($\alpha_t \times 10^{-3}$) and anisotropy, $\Delta\alpha_t$ (Eq. 2), of a typical (VCA) atom in each $Fe_{50}Co_{50}$(001) bcc surface layer for magnetization along [010] and [110] directions. In each line, the sum of all $\alpha_{ij}$ in the same layer is considered. Outermost (layer 1) and deeper layers (2-5).

| Layer | $\alpha_t$ [010] | $\alpha_t$ [110] | $\Delta\alpha_t$ |
|---|---|---|---|
| 1 | 7.00 | 14.17 | +102.4% |
| 2 | 1.28 | 1.16 | −9.4% |
| 3 | 2.83 | 3.30 | +16.6% |
| 4 | 2.18 | 1.99 | −8.7% |
| 5 | 2.54 | 2.53 | −0.4% |

We also studied the impact of bct-like distortions *in* the surface, initially by considering the VCA model. Similar to the bulk case, tetragonal distortions may be important for the damping anisotropy at the surface, e.g. when local structural defects are present. Therefore, localized bct-like distortions of the VCA medium in the surface, particularly involving the most external layer were investigated. The structural model was similar to what was used for the $Fe_{50}Co_{50}$ bulk, considering $\frac{c}{a} = 1.09$ (see SM). Our calculations show that tetragonal relaxations around a typical site in the surface induce a $\Delta\alpha_t \sim 75\%$, from $\alpha_t^{[010]} = 8.94 \times 10^{-3}$ to $\alpha_t^{[110]} = 15.68 \times 10^{-3}$. The main effect of these distortions is an enhancement of the absolute damping values in each direction with respect to the pristine (bcc) system. This is due to an increase on $\alpha_{onsite}$, from $\alpha_{onsite}^{[010]} = 7.4 \times 10^{-3}$ to $\alpha_{onsite}^{[010]} = 9.5 \times 10^{-3}$, and from $\alpha_{onsite}^{[110]} = 8.7 \times 10^{-3}$ to $\alpha_{onsite}^{[110]} = 11.7 \times 10^{-3}$; the resulting non-local contributions remains similar to the undistorted case. The influence of bct-like distortions on the large damping value in the $Fe_{50}Co_{50}$ surface is in line with results of Mandal *et al.* [30], and is related to the transition of minority spin electrons around $E_F$.

We then considered explicit 10-atom $Fe_{50}Co_{50}$ clusters embedded in a VCA FeCo surface matrix. The results from these calculations were obtained as an average over 16 Fe-centered and 16 Co-centered clusters. We considered clusters with undistorted bcc crystal structure (Fig. 2, yellow open circles) as well as clusters with local tetragonal distortions (Fig. 2, black open circles). As shown in Fig. 2 the explicit local tetragonal distortion influences the damping values ($\alpha_t^{[010]} = 10.03 \times 10^{-3}$ and $\alpha_t^{[110]} = 14.86 \times 10^{-3}$) and the anisotropy, but not enough to reproduce the huge values reported in the experiments.

A summary of the results obtained for each undistorted FeCo cluster at the surface is shown in Fig. 3: Co-centered clusters in Fig. 3(a) and Fe-centered clusters in Fig. 3(b). A large variation of $\alpha_t$ values is seen from cluster to cluster, depending on the spatial distribution of atomic species. It is clear that, $\alpha_t$ is larger when there is a larger number of Fe atoms in the surface layer that surrounds the central, reference cluster site. This correlation can be seen by the numbers in parenthesis on top of the blue symbols (total damping for each of the 16 clusters that were considered) in Fig. 3. We also notice from the figure that the damping in Fe-centered clusters are lower than in Co-centered, and that the [010] magnetization direction exhibit always lower values. In the *Inset* of Fig. 3 the onsite contributions to the damping, $\alpha_{onsite}$, and the LDOS at $E_F$ in the central site of each cluster are shown: a correlation, where both trends are the same, can be observed. The results in Fig. 3 shows that the neighbourhood influences not only the local electronic structure at the reference site (changing $n(E_F)$ and $\alpha_{onsite}$), but also modifies the non-local damping $\alpha_{ij}$, leading to the calculated $\alpha_t$. In other words, the local spatial distribution affects how the total damping is manifested, something which is expressed differently among different clusters. This may open up for materials engineering of local and non-local contributions to the damping.

*Conclusions:* We demonstrate here that real-space electronic structure, based on density functional theory, yield a large Gilbert damping anisotropy in $Fe_{50}Co_{50}$ alloys. Theory leads to a large damping anisotropy, when the magnetization changes from the [010] to the [110] direction, which can be as high as $\sim 100\%$ (or 200% in the minimum-maximum damping ratio) when surface calculations are considered. This is in particular found for contributions from surface atoms in the outermost layer. Hence the results presented here represents one more example, in addition to the well known enhanced surface orbital moment [31], of the so-called interfacial spin-orbit coupling. This damping anisotropy, which holds a bcc-like fourfold ($C_{4v}$) symmetry, has a close relation to the LDOS difference of the most external layer at $E_F$ (majorly contributed by the minority $d$ states), as well as to the orbital moment anisotropy with a 90° phase. As a distinct example of an interface, we consider explicitly the $Fe_{50}Co_{50}$ cluster description of the alloy. In this case, besides an onsite contribution, we find that the damping anisotropy is mostly influenced by non-local next-nearest-neighbours interactions.

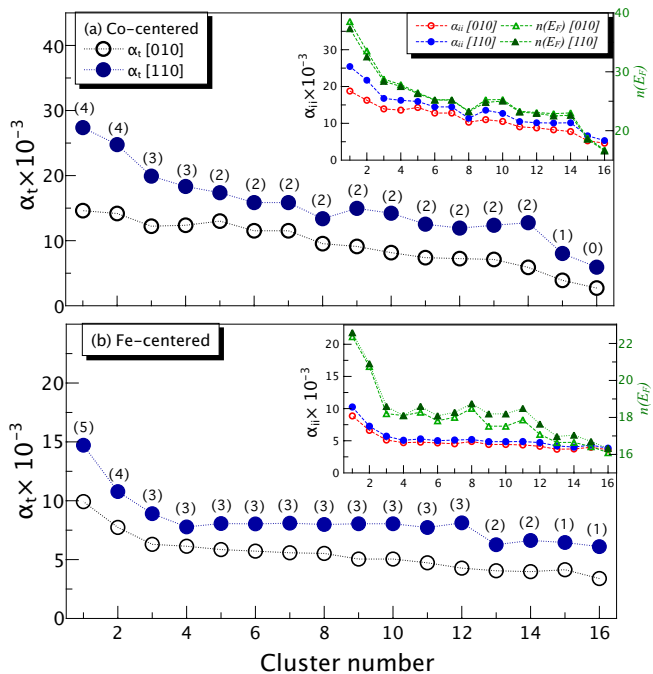

Figure 3. Damping for the [010] (open circles) and [110] (full circles) magnetization directions for distinct types of 10-atom $Fe_{50}Co_{50}$ bcc clusters, embedded in VCA $Fe_{50}Co_{50}$(001) bcc and without any distortion around the reference atom (for which $\alpha_t$ and $\alpha_{onsite}$ are shown). (a) Co-centered and (b) Fe-centered clusters. The quantity of Fe atoms in the surface layers (near vacuum) are indicated by the numbers in parenthesis and the results have been ordered such that larger values are to the left in the plots. *Insets*: $\alpha_{onsite}$ for the [010] (red open circles) and [110] (blue filled circles) magnetization directions, and corresponding local density of states, $n(E_F)$, at the Fermi level (green filled and unfilled triangles) at the central atom (placed in the outermost layer) for both types of clusters. Lines are guides for eyes.

Several Gilbert damping anisotropy origins are also demonstrated here, primarily related to the presence of interfaces, alloy composition and local structural distortions (as summarized in Table S6, in the SM [32]). Primarily we find that: (*i*) the presence of Co introduces an enhanced spin-orbit interaction and can locally modify the non-local damping terms; (*ii*) the randomness of Co in the material, can modestly increase $\Delta\alpha_t$ as a total effect by creating Co-concentrated clusters with enhanced damping; (*iii*) at the surface, the spatial distribution of Fe/Co, increases the damping when more Fe atoms are present in the outermost layer; and (*iv*) the existence of local, tetragonal distortions, which act in favour (via SOC) of the absolute damping enhancement, by modifying the $\alpha_{onsite}$ of the reference atom, and could locally change the spin relaxation time. Furthermore, in relationship to the work in Ref. [10], we show here that bulk like tetragonal distortions, that in Ref. [10] were suggested to be the key reason behind the observed huge anisotropy of the damping, can in fact not explain the experimental data. Such distortions were explicitly considered here, using state-of-the-art theory, and we clearly demonstrate that this alone can not account for the observations.

Although having a similar trend as the experimental results of Ref. [10], we do not reproduce the most extreme maximum-minimum ratio reported in the experiment, $\sim 400\%$ (or $\Delta\alpha_t \sim 300\%$). The measured damping does however include effects beyond the intrinsic damping that is calculated from our electronic structure methodology. Other mechanisms are known to influence the damping parameter, such as contributions from eddy currents, spin-pumping, and magnon scattering, to name a few. Thus it is possible that a significant part of the measured anisotropy is caused by other, extrinsic, mechanisms. Despite reasons for differences between observation and experiment on films of $Fe_{50}Co_{50}$ alloys, the advancements presented here provide new insights on the intrinsic damping anisotropy mechanisms, something which is relevant for the design of new magnetic devices.

*Acknowledgements:* H.M.P. and A.B.K. acknowledge financial support from CAPES, CNPq and FAPESP, Brazil. The calculations were performed at the computational facilities of the HPC-USP/CENAPAD-UNICAMP (Brazil), at the National Laboratory for Scientific Computing (LNCC/MCTI, Brazil), and at the Swedish National Infrastructure for Computing (SNIC). I.M. acknowledge financial support from CAPES, Finance Code 001, process n° 88882.332894/2018-01, and in the Institutional Program of Overseas Sandwich Doctorate, process n° 88881.187258/2018-01. O.E. acknowledges support from the Knut och Alice Wallenberg (KAW) foundation, the Swedish research council (VR), the Foundation for Strategic Research (SSF), the Swedish energy agency (Energimyndigheten), eSSENCE, STandUPP, and the ERC (synergy grant FASTCORR). D.T. acknowledges support from the Swedish Research Council (VR) through Grant No. 2019-03666.

# Supplemental Material to "Mechanisms behind large Gilbert damping anisotropies"

I. P. Miranda[1], A. B. Klautau[2],[*] A. Bergman[3], D. Thonig[3,4], H. M. Petrilli[1], and O. Eriksson[3,4]

[1] *Universidade de São Paulo, Instituto de Física,*
*Rua do Matão, 1371, 05508-090, São Paulo, SP, Brazil*
[2] *Faculdade de Física, Universidade Federal do Pará, Belém, PA, Brazil*
[3] *Department of Physics and Astronomy, Uppsala University, Box 516, SE-75120 Uppsala, Sweden and*
[4] *School of Science and Technology, Örebro University, Fakultetsgatan 1, SE-701 82 Örebro, Sweden*

(Dated: November 22, 2021)

## I. Theory

The torque-correlation model, first introduced by Kamberský [1], and later elaborated by Gilmore *et al.* [2], can be considered as both a generalization and an extended version of the breathing Fermi surface model, which relates the damping of the electronic spin orientation, with the variation in the Fermi surface when the local magnetic moment is changed. In this scenario, and considering the collinear limit of the magnetic ordering, due to the spin-orbit coupling (SOC), the tilting in magnetization $\hat{\mathbf{m}}$ by a small change $\delta\hat{\mathbf{m}}$ generates a non-equilibrium population state which relaxes within a time $\tau$ towards the equilibrium. We use an angle $\theta$, to represent the rotation of the magnetization direction $\delta\hat{\mathbf{m}}$. If the Bloch states of the systems are characterized by the generic band index $n$ at wavevector $\mathbf{k}$ (with energies $\epsilon_{\mathbf{k},n}$), it is possible to define a *tensor* for the damping, that has matrix elements (adopting the isotropic relaxation time approximation)

$$\alpha^{\nu\mu} = \frac{g\pi}{m}\sum_{n,m}\frac{d\mathbf{k}}{(2\pi)^3}\eta(\epsilon_{\mathbf{k},n})\left(\frac{\partial\epsilon_{\mathbf{k},n}}{\partial\theta}\right)_\nu\left(\frac{\partial\epsilon_{\mathbf{k},m}}{\partial\theta}\right)_\mu\frac{\tau}{\hbar} \quad (\text{S1})$$

which accounts for both intraband ($n = m$, conductivity-like) and interband ($n \neq m$, resistivity-like) contributions [2]. Here $\mu, \nu$ are Cartesian coordinate indices, that will be described in more detail in the discussion below, while $\eta(\epsilon_{\mathbf{k},n}) = \left.\frac{\partial f(\epsilon)}{\partial \epsilon}\right|_{\epsilon_{\mathbf{k},n}}$ is the derivative of the Fermi distribution, $f$, with respect to the energy $\epsilon$, and $n, m$ are band indices. Therefore, the torque-correlation model correlates the spin damping to variations of the energy of single-particle states with respect to the variation of the spin direction $\theta$, i.e. $\frac{\partial \epsilon_{\mathbf{k},n}}{\partial\theta}$. Using the Hellmann-Feynmann theorem, which states that $\frac{\partial \epsilon_{\mathbf{k},n}}{\partial\theta} = \langle\psi_{\mathbf{k},n}|\frac{\partial\mathcal{H}}{\partial\theta}|\psi_{\mathbf{k},n}\rangle$, and the fact that only the spin-orbit Hamiltonian $\mathcal{H}_{so}$ changes with the magnetization direction, the spin-orbit energy variation is given by

$$\frac{\partial\epsilon_{\mathbf{k},n}(\theta)}{\partial\theta} = \langle\psi_{\mathbf{k},n}|\frac{\partial}{\partial\theta}\left(e^{i\boldsymbol{\sigma}\cdot\hat{\mathbf{n}}\theta}\mathcal{H}_{so}e^{-i\boldsymbol{\sigma}\cdot\hat{\mathbf{n}}\theta}\right)|\psi_{\mathbf{k},n}\rangle \quad (\text{S2})$$

in which $\boldsymbol{\sigma}$ represents the Pauli matrices vector, and $\hat{\mathbf{n}}$ is the direction around which the local moment has been rotated. The expression in Eq. S2 can be easily transformed into $\frac{\partial\epsilon_{\mathbf{k},n}(\theta)}{\partial\theta} = i\langle\psi_{\mathbf{k},n}|[\boldsymbol{\sigma}\cdot\hat{\mathbf{n}},\mathcal{H}_{so}]|\psi_{\mathbf{k},n}\rangle$ and we call $\hat{T} = [\boldsymbol{\sigma}\cdot\hat{\mathbf{n}},\mathcal{H}_{so}]$ the *torque* operator. In view of this, it is straightforward that, in the collinear case in which all spins are aligned to the $z$ direction, $\boldsymbol{\sigma}\cdot\hat{\mathbf{n}} = \sigma^\mu$ ($\mu = x,y,z$), originating the simplest $\{x,y,z\}$-dependent torque operator $\hat{T}^\mu$. Putting together the information on Eqs. S1 and S2, and using the fact that the imaginary part of the Greens' functions can be expressed, in Lehmann representation, as $\text{Im}\hat{G}(\epsilon\pm i\Lambda) = -\frac{1}{\pi}\sum_n\frac{\Lambda}{(\epsilon-\epsilon_n)^2+\Lambda^2}|n\rangle\langle n|$, then it is possible to write in reciprocal-space [3]:

$$\alpha^{\nu\mu} = \frac{g}{m\pi}\int\int\eta(\epsilon)\text{Tr}\left(\hat{T}^\nu\text{Im}\hat{G}\hat{T}^\mu\text{Im}\hat{G}\right)d\epsilon\frac{d\mathbf{k}}{(2\pi)^3}. \quad (\text{S3})$$

In a real-space formalism, the Fourier transformation of the Green's function is used to find a very similar expression emerges for the damping element $\alpha_{ij}^{\nu\mu}$ relative to two atomic sites $i$ and $j$ (at positions $\mathbf{r}_i$ and $\mathbf{r}_j$, respectively) in the material:

$$\alpha_{ij}^{\nu\mu} = \frac{g}{m_i\pi}\int\eta(\epsilon)\text{Tr}\left(\hat{T}_i^\nu\text{Im}\hat{G}_{ij}\hat{T}_j^\mu\text{Im}\hat{G}_{ji}\right)d\epsilon, \quad (\text{S4})$$

where we defined $m_i = (m_{orb} + m_{spin})$ as the total magnetic moment localized in the reference atomic site $i$ in the pair $\{i,j\}$. The electron temperature that enters into $\eta(\varepsilon)$ is zero and, consequently, the energy integral is performed only at the Fermi energy. In this formalism, then, the intraband and interband terms are replaced by onsite ($i = j$) and non-local ($i \neq j$) terms. After calculation of all components of Eq. S4 in a collinear magnetic background, we get a tensor of the form

$$\boldsymbol{\alpha}_{ij} = \begin{pmatrix} \alpha^{xx} & \alpha^{xy} & \alpha^{xz} \\ \alpha^{yx} & \alpha^{yy} & \alpha^{yz} \\ \alpha^{zx} & \alpha^{zy} & \alpha^{zz} \end{pmatrix}, \quad (\text{S5})$$

which can be used in the generalized atomistic Landau-Lifshitz-Gilbert (LLG) equation for the spin-dynamics of magnetic moment on site $i$ [4]: $\frac{\partial\mathbf{m}_i}{\partial t} = \mathbf{m}_i \times \left(-\gamma\mathbf{B}_i^{\text{eff}} + \sum_j\frac{\boldsymbol{\alpha}_{ij}}{m_j}\cdot\frac{\partial\mathbf{m}_j}{\partial t}\right)$. Supposing that all spins are parallel to the local $z$ direction, we can define the scalar $\alpha$ value as the average between components $\alpha^{xx}$ and $\alpha^{yy}$, that is: $\alpha = \frac{1}{2}(\alpha^{xx} + \alpha^{yy})$.

Once one has calculated the onsite ($\alpha_{onsite}$) and the non-local ($\alpha_{ij}$) damping parameters with respect to the site of interest $i$, the total value, $\alpha_t$, can be defined as the sum of all these $\alpha$'s:

$$\alpha_t = \sum_{\{i,j\}} \alpha_{ij}. \tag{S6}$$

In order to obtain the total damping in an heterogeneous atomic system (more than one element type), such as $Fe_{50}Co_{50}$ (with explicit Fe/Co atoms), we consider the weighted average between the different total local damping values ($\alpha_t^i$), namely:

$$\alpha_t = \frac{1}{M^{\text{eff}}} \sum_i m_i \alpha_t^i, \tag{S7}$$

where $m_i$ is the local magnetic moment at site $i$, and $M^{\text{eff}} = \sum_i m_i$ is the summed total effective magnetization. This equation is based on the fact that, in FMR experiments, the magnetic moments are excited in a zone-centered, collective mode (Kittel mode). In the results presented here, Eq. S7 was used to calculate $\alpha_t$ of $B2$-FeCo, both in bcc and bct structures.

**II. Details of calculations**

The real-space linear muffin-tin orbital on the atomic-sphere approximation (RS-LMTO-ASA) [5] is a well-established method in the framework of the DFT to describe the electronic structure of metallic bulks [6, 7], surfaces [8, 9] and particularly embedded [10] or absorbed [11–14] finite cluster systems. The RS-LMTO-ASA is based on the LMTO-ASA formalism [15], and uses the recursion method [16] to solve the eigenvalue problem directly in real-space. This feature makes the method suitable for the calculation of local properties, since it does not depend on translational symmetry.

The calculations performed here are fully self-consistent, and the spin densities were treated within the local spin-density approximation (LSDA) [17]. In all cases, we considered the spin-orbit coupling as a $l \cdot s$ term included in each variational step [18–20]. The spin-orbit is strictly necessary for the damping calculations due to its strong dependence on the torque operators, $\hat{T}$. In the recursion method, the continued fractions have been truncated with the Beer-Pettifor terminator [21] after 22 recursion levels ($LL = 22$). The imaginary part that comes from the terminator was considered as a natural choice for the broadening $\Lambda$ to build the Green's functions $\hat{G}(\epsilon + i\Lambda)$, which led to reliable $\alpha$ parameters in comparison with previous results (see Table S1).

To account for the Co randomness in the experimental $Fe_{50}Co_{50}$ films [22], some systems were modeled in terms of the virtual crystal approximation (VCA) medium of $Fe_{50}Co_{50}$, considering the bcc (or the bct) matrix to have the same number of valence electrons as $Fe_{50}Co_{50}$ (8.5 $e^-$). However, we also investigated the role of the Co presence, as well as the influence of its randomness (or ordering), by simulating the $B2$ (CsCl) FeCo structure ($a = a_{Fe}$). The VCA $Fe_{50}Co_{50}$ and $B2$-FeCo bulks were simulated by a large matrix containing 8393 atoms in real-space, the first generated by using the Fe bcc lattice parameter ($a_{Fe} = 2.87$ Å) and the latter using the optimized lattice parameter ($a = 2.84$ Å). This $a$ choice in VCA $Fe_{50}Co_{50}$ was based on the fact that it is easier to compare damping results for $Fe_{50}Co_{50}$ alloy and pure Fe bcc bulk if the lattice parameters are the same, and the use of the $a_{Fe}$ has shown to produce trustworthy $\alpha_t$ values. On the other hand, bct bulk structures with $\frac{c}{a} = 1.15$ ($B2$-FeCo bct and VCA $Fe_{50}Co_{50}$ bct) are based on even larger matrices containing 49412 atoms. The respective surfaces were simulated by semi-matrices of the same kind (4488 and 19700 atoms, respectively), considering one layer of empty spheres above the outermost $Fe_{50}Co_{50}$ (or pure Fe) layer, in order to provide a basis for the wave functions in the vacuum and to treat the charge transfers correctly.

We notice that the investigations presented here are based on a (001)-oriented $Fe_{50}Co_{50}$ film, in which only a small lattice relaxation normal to the surface is expected to occur ($\sim 0.1\%$ [23]).

Damping parameters of Fe-centered and Co-centered clusters, embedded in an $Fe_{50}Co_{50}$ VCA medium, have been calculated (explicitly) site by site. In all cases, these defects are treated self-consistently, and the potential parameters of the remaining sites were fixed at bulk/pristine VCA surface values, according to its environment. When inside the bulk, we placed the central (reference) atom of the cell in a typical site far away from the faces of the real-space matrix, avoiding any unwanted surface effects. We considered as impurities the nearest 14 atoms (first and second nearest neighbours, up to $1a$) from the central atom, treating also this sites self-consistently, in a total of 15 atoms. We calculated 10 cases with Fe and Co atoms randomly positioned: 5 with Fe as the central atom (Fe-centered) and 5 with Co as the central atom (Co-centered). An example (namely cluster #1 of Tables S3 and S4), of one of these clusters embedded in bulk, is represented in Fig. S1(a). As the self-consistent clusters have always a total of 15 atoms, the Fe (Co) concentration is about 47% (53%) or vice-versa. On the other hand, when inside the surface, we placed the central (reference) atom of the cluster in a typical site of the most external layer (near vacuum), since this has shown to be the layer where the damping anisotropy is larger. Therefore, we considered as impurities the reference atom itself and the nearest 9 atoms (up to $1a$), in a total of 10 atoms (and giving a perfect 50% (50%) concentration). An example of one of these clusters embedded in a surface is shown in Fig. S1(b).

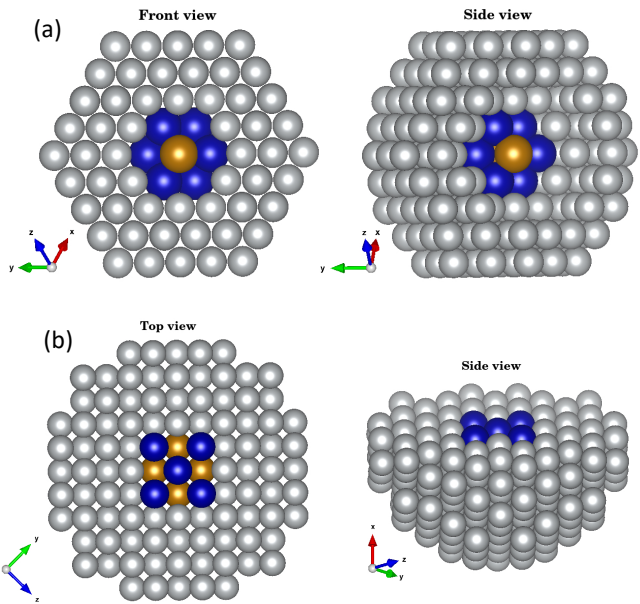

Figure S1. (Color online) Schematic representation of an example of: (a) Fe-centered 15-atom cluster embedded in a VCA $Fe_{50}Co_{50}$ bcc bulk medium; (b) Co-centered 10-atom cluster embedded in a VCA $Fe_{50}Co_{50}$(001) bcc surface medium. Yellow and blue spheres represent Fe and Co atoms, respectively, while gray atoms represent the VCA $Fe_{50}Co_{50}$ sites (8.5 valence $e^-$). The Fe(Co) concentration in the clusters are: (a) 53% (47%) and (b) 50% (50%). The total number of atoms including the surrounding VCA sites are: (a) 339 and (b) 293. They were all accounted in the sum to obtain $\alpha_t$ at the central (reference) Fe (a) and Co (b) site.

To simulate a bct-like bulk distortion, the 8 first neighbours of the central atom were stretched in the $c$ direction, resulting in a $\frac{c}{a} = 1.15$ ratio. On the other hand, when embedded on the $Fe_{50}Co_{50}$(001) bcc surface, the central (reference) atom is placed in the outermost layer (near vacuum), and we simulate a bct distortion by stretching the 4 nearest-neighbours (on the second layer) to reproduce a $\frac{c}{a} = 1.09$ ratio (the maximum percentage that the atoms, in these conditions, could be moved to form a bct-like defect). In this case, a total of 10 atoms (the nearest 9 atoms from the central one – up to $1a$ – and the reference atom) were treated self-consistently, analogous to as shown in Fig. S1(b). As in the case of the pristine bcc $Fe_{50}Co_{50}$ clusters embedded in the VCA surface, we considered a total of 32 10-atom clusters with different Fe/Co spatial distributions, being 16 Fe-centered, and 16 Co-centered.

## III. Comparison with previous results

The *ab-initio* calculation of the Gilbert damping, in the collinear limit, is not a new feature in the literature. Mainly, the reported theoretical damping results are for bulk systems [2, 4, 24–28], but, some of them even studied free surfaces [29]. Therefore, in order to demonstrate the reliability of the on-site and total damping calculations implemented here in real-space, a comparison of the presently obtained with previous (experimental and theoretical) results, are shown in Table S1. As can be seen, our results show a good agreement with previously obtained $\alpha$ values, including some important trends already predicted before. For example, the reduced Gilbert damping of Co hcp with respect to the Co fcc due to the reduction of the density of states at the Fermi level [24, 28], ($\sim 10.92$ states/Ry-atom in the hcp case and $\sim 16.14$ states/Ry-atom in the fcc case).

## IV. Details of the calculated damping values

The damping values obtained for the systems studied here are shown in Tables (S2-S5). These data can be useful for the full understanding of the results presented in the main text. For easy reference, in Table S2 the $\alpha_t$ of a typical atom in each system (bulk or surface) for different spin quantization axes are shown. These data are plotted in Fig. 2 of the main text. The obtained values show that, indeed, for bulk systems the damping anisotropies are not so pronounced as in the case of $Fe_{50}Co_{50}$(001) bcc surface.

As observed in Table S2, the increase in $\alpha_t$ when changing from the bcc $Fe_{50}Co_{50}$ ($\frac{c}{a} = 1$) to the bct $Fe_{50}Co_{50}$ bulk structure ($\frac{c}{a} = 1.15$) is qualitatively consistent to what was obtained by Mandal *et al.* [33] (from $\alpha_t = 6.6 \times 10^{-3}$ in the bcc to $\alpha_t = 17.8 \times 10^{-3}$ in the bct, with $\frac{c}{a} = 1.33$ [33]).

Tables S3 and S4 refer to the damping anisotropies ($\Delta\alpha_t$) for all Fe-centered and Co-centered clusters studied here, with different approaches: ($i$) bcc clusters embedded in the VCA medium (Table S3) and ($ii$) bct-like clusters embedded in the VCA medium (Table S4).

In comparison with bct-like clusters, we found larger absolute $\alpha_t$ values but lower damping anisotropies. In all cases, Fe-centered clusters present higher $\Delta\alpha_t$ percentages.

In Table S5 the onsite damping anisotropies ($\Delta\alpha_{onsite}$) for each layer of the $Fe_{50}Co_{50}$(001) bcc surface ("1" represents the layer closest to vacuum) are shown. In comparison with the total damping anisotropies (Table I of the main text), much lower percentages are found, demonstrating that the damping anisotropy effect comes majorly from the non-local damping contributions.

The most important results concerning the largest damping anisotropies are summarized in Table S6, below.

Table S1. Total damping values ($\times 10^{-3}$) calculated for some bulk and surface systems, and the comparison with previous literature results. The onsite contributions are indicated between parentheses, while the total damping, $\alpha_t$, are indicated without any symbols. All values were obtained considering the [001] magnetization axis. The VCA was adopted for alloys, except for the $Fe_{50}Co_{50}$ bcc in the $B2$ structure (see Eq. S7). Also shown the broadening $\Lambda$ value considered in the calculations.

| **Bulks** | $a$ (Å) | **This work** | **Theoretical** | **Experimental** | $\Lambda$ (eV) |
|---|---|---|---|---|---|
| Fe bcc | 2.87 | 4.2(1.6) | 1.3 [2][a]/(3.6) [4] | 1.9 [30]/2.2 [31] | |
| $Fe_{70}Co_{30}$ bcc | 2.87 | 2.5(0.7) | – | 3 − 5 [32][d] | |
| $Fe_{50}Co_{50}$ bcc | 2.87 | 3.7(1.0) [VCA]/2.3(1.0) [$B2$] | 1.0 [25][c][VCA]/6.6 [33] [$B2$] | 2.3 [27] | |
| Ni fcc | 3.52 | 27.8(57.7) | 23.7 [34]/(21.6 [4])[b] | 26.0 [31]/24.0 [35] | |
| $Ni_{80}Fe_{20}$ (Py) fcc | 3.52 | 9.8(12.1) | 3.9 [25][c] | 8.0 [30]/5.0 [35] | $\sim 5 \times 10^{-2}$ |
| Co fcc | 3.61 [3] | 3.2(5.3) | 5.7 [28]/(3.9 [4])[b] | 11.0 [30] | |
| Co hcp | 2.48/4.04 [28] | 2.1(6.2) | 3.0 [28] | 3.7 [31] | |
| $Co_{85}Mn_{15}$ bcc [36] | 2.87 [28] | 6.2(4.2) | 6.6 [28] | – | |
| $Co_{90}Fe_{10}$ fcc | 3.56 [37] | 3.6(4.2) | – | 3.0 [35]/4.8 [37] | |
| **Surfaces** | $a$ (Å) | **This work** | **Theoretical** | **Experimental** | |
| Fe(001) bcc [110] | 2.87 | 5.8(5.4)[e] | – | 7.2 [38][h]/6.5[39][i] | |
| Fe(001) bcc [100] | 2.87 | 3.9(4.4)[f] | $\sim 4$ [29][g] | 4.2 [40][j] | |
| Ni(001) fcc | 3.52 | 80.0(129.6) | $\sim 10$ [29][g]/12.7 [41][m] | 22.1 [42][l] | |
| PdFe/Ir(111) [43] fcc | 3.84 | 3.9(2.7)[n] | – | – | |
| PdCo/Ir(111) [44] fcc | 3.84 | 3.2(14.7)[o] | – | – | |

[a] With $\Lambda \sim 2 \times 10^{-2}$ eV.
[b] With $\Lambda = 5 \times 10^{-3}$ eV.
[c] With $\Lambda \sim 1.4 \times 10^{-4}$ eV.
[d] For a 28% Co concentration, but the results do not significantly change for a 30% Co concentration. Range including results before and after annealing.
[e] Of a typical atom in the more external surface layer (in contact with vacuum), in the [110] magnetization direction.
[f] Of a typical atom in the more external surface layer (in contact with vacuum), in the [100] magnetization direction.
[g] For a (001) bcc surface with thickness of $N = 8$ ML (the same number of slabs as in our calculations), and $\Lambda = 10^{-2}$ eV.
[h] Anisotropic damping obtained for a 0.9 nm Fe/GaAs(001) thin film (sample S2 in Ref. [38]) in the [110] magnetization direction.
[i] Anisotropic damping obtained for a 1.14 nm Fe/InAs(001) thin film in the in-plane [110] hard magnetization axis.
[j] For a 25-nm-thick Fe films grown on MgO(001).
[k] For epitaxial Fe(001) films grown on GaAs(001) and covered by Au, Pd, and Cr capping layers.
[l] Intrinsic Gilbert damping for a free 4×[Co(0.2 nm)/Ni(0.6 nm)](111) multilayer. Not the same system as Ni(001), but the nearest system found in literature.
[m] For a Co | Ni multilayer with Ni thickness of 4 ML (fcc stacking).
[n] Of a typical atom in the Fe layer.
[o] Of a typical atom in the Co layer.

The alloys with short-range orders (SRO) are described as FeCo clusters (with explicit Fe and Co atoms) embedded in the $Fe_{50}Co_{50}$ VCA medium – with and without the bct-like distortion. In this case, the damping is calculated as a weighted average (Eq. S7). As discussed in the main text, it can be seen from Table S6 that distortions and disorder can increase the anisotropy but the major effect comes from the surface. We notice that the number of clusters considered is limited in the statistical average.

### IV. Kambersky's simplified formula

In order to connect the anisotropy of the Gilbert damping to features in the electronic structure, we consider in the following Kambersky's simplified formula for Gilbert damping [47, 48]

$$\alpha = \frac{1}{\gamma M_s} \left(\frac{\gamma}{2}\right)^2 n(E_F) \xi^2 \frac{(g-2)^2}{\tau} . \quad (S8)$$

Here, $\gamma$ is the gyromagnetic ratio, $n(E_F)$ represents the LDOS at the Fermi level, $\xi$ is the SOC strength, $\tau$ is the electron scattering time, $M_s$ is the spin magnetic moment, and $g$ is the spectroscopic $g$-factor [35, 49]. Note that Eq. S8 demonstrates the direct relation between $\alpha$ and $n(E_F)$, often discussed in the literature, e.g., in Ref. [27]. Our first principles calculations have shown no significant change in $\xi$, upon variation of the magnetization axis, for the FeCo systems ($\xi_{Co} = 71.02$ meV and $\xi_{Fe} = 53.47$ meV). Hence, we can soundly relate the damping anisotropy $\Delta \alpha_t$ to $\Delta n(E_F)$.

Figure S2 shows how the LDOS difference (per atom) $\Delta n(E)$ between the [010] and [110] magnetization directions is developed in pure Fe(001) bcc and in VCA $Fe_{50}Co_{50}$(001) bcc surfaces, respectively. In both cases,

Table S2. Total damping ($\alpha_t \times 10^{-3}$) of a typical atom in each system for the spin quantization axes [010] ($\theta_H = 90°$) and [110] ($\theta_H = 45°$); also shown for the [001] and [111]. Bulk and surface bct systems are simulated with $\frac{c}{a} = 1.15$.

**Bulks**

| Bulk | $\alpha_t$ [010] | $\alpha_t$ [110] | $\Delta\alpha_t$ |
|---|---|---|---|
| Fe bcc | 4.18 | 4.31 | +3.1%[a] |
| $B2$-FeCo bcc | 2.28 | 2.44 | +7.2% |
| $B2$-FeCo bct | 7.76 | 8.85 | +12.4% |
| VCA Fe$_{50}$Co$_{50}$ bcc | 3.70 | 4.18 | +13.0% |
| VCA Fe$_{50}$Co$_{50}$ bct | 4.69 | 5.10 | +8.7% |
| | $\alpha_t$ [010] | $\alpha_t$ [001] | $\Delta\alpha_t$ |
| $B2$-FeCo bct | 7.76 | 10.21 | +24.1% |
| VCA Fe$_{50}$Co$_{50}$ bct | 4.69 | 5.75 | +22.6% |
| | $\alpha_t$ [010] | $\alpha_t$ [111] | $\Delta\alpha_t$ |
| Fe bcc | 4.18 | 4.56 | +9.1%[b] |

**Surfaces**

| Surface | $\alpha_t$ [010] | $\alpha_t$ [110] | $\Delta\alpha_t$ |
|---|---|---|---|
| Fe(001) bcc | 3.85 | 5.75 | +49.4% |
| Fe/GaAs(001) bcc [38] | 4.7(7) | 7.2(7) | +53(27)%[c] |
| Fe/MgO(001) bcc [45] | 3.20(25) | 6.15(20) | +92(14)%[d] |
| VCA Fe$_{50}$Co$_{50}$(001) bcc | 7.00 | 14.17 | +102.4% |
| VCA Fe$_{50}$Co$_{50}$(001) bct | 15.20 | 14.80 | −2.6% |
| | $\alpha_t$ [010] | $\alpha_t$ [001] | $\Delta\alpha_t$ |
| VCA Fe$_{50}$Co$_{50}$(001) bct | 15.20 | 15.56 | +2.4% |
| VCA Fe$_{50}$Co$_{50}$(001) bcc | 7.00 | 9.85 | +40.7% |

[a] Mankovsky *et al.* [24] find a damping anisotropy of ∼ 12% for bulk Fe bcc at low temperatures (∼ 50 K) between [010] and [011] magnetization directions. For this result, the definition $\alpha = \frac{1}{2}(\alpha^{xx} + \alpha^{yy})$ was used.
[b] This result agrees with Gilmore *et al.* [46], which find that the total damping of pure Fe bcc presents its higher value in the [111] crystallographic orientation and the lower value in the [001] direction, except for high scattering rates. Also agrees with Mankovsky *et al.* [24] results.
[c] Anisotropic damping obtained for a 0.9 nm Fe/GaAs(001) thin film (sample S2 in Ref. [38]) in the [010] and [110] magnetization directions.
[d] For a Fe(15 nm)/MgO(001) film at $T = 4.5$ K in the highest applied magnetic field, in which only intrinsic contributions to the anisotropic damping are left.

the chosen layer, denoted as *first*, is the most external one (near vacuum). the VCA Fe$_{50}$Co$_{50}$(001) bcc we also calculated $\Delta n(E)$ for all layers summed (total DOS difference).

As can be seen, although in all cases the quantity $\Delta n(E)$ exhibits some oscillations, differently from what we observe for the pure Fe(001) surface case, at the Fermi energy, there is a non-negligible difference in the minority spin channel (3$d$ states) for the VCA Fe$_{50}$Co$_{50}$(001). Considering the results presented in Table I (main text) the larger contribution to the damping anisotropy comes from the most external layer. The results by Li *et al.* [22] indicate a small difference (for two magnetization directions) of the total density of states at the Fermi

Table S3. Total damping anisotropy ($\times 10^{-3}$) of all studied Co-centered and Fe-centered bcc clusters for the spin-quantization axis [010] and [110], considering the 15-atom FeCo cluster together with the VCA medium in the summation for total damping.

| Co-centered | | | |
|---|---|---|---|
| Cluster # | $\alpha_t$ [010] | $\alpha_t$ [110] | $\Delta\alpha_t$ |
| 1 | 10.11 | 9.65 | 4.8% |
| 2 | 8.09 | 6.96 | 16.2% |
| 3 | 7.81 | 7.02 | 11.3% |
| 4 | 7.11 | 7.02 | 1.3% |
| 5 | 7.48 | 6.88 | 8.7% |
| **Average** | 8.12 | 7.51 | 8.1% |
| Fe-centered | | | |
| Cluster # | $\alpha_t$ [010] | $\alpha_t$ [110] | $\Delta\alpha_t$ |
| 1 | 2.68 | 2.03 | 32.0% |
| 2 | 2.49 | 2.05 | 21.5% |
| 3 | 2.56 | 1.86 | 37.6% |
| 4 | 2.45 | 1.79 | 36.9% |
| 5 | 2.76 | 2.01 | 37.3% |
| **Average** | 2.59 | 1.95 | 32.8% |

Table S4. Total damping anisotropy ($\times 10^{-3}$) of all studied Co-centered and Fe-centered bcc clusters for the spin quantization axis [010] and [110], with bct-like distortions ($\frac{c}{a} = 1.15$), considering the 15-atom FeCo cluster together with the VCA medium in the summation for total damping.

| Co-centered | | | |
|---|---|---|---|
| Cluster # | $\alpha_t$ [010] | $\alpha_t$ [110] | $\Delta\alpha_t$ |
| 1 | 5.85 | 4.37 | 33.9% |
| 2 | 5.95 | 4.21 | 41.3% |
| 3 | 5.88 | 4.35 | 35.2% |
| 4 | 5.90 | 4.41 | 33.8% |
| 5 | 5.86 | 4.34 | 35.0% |
| **Average** | 5.89 | 4.34 | 35.7% |
| Fe-centered | | | |
| Cluster # | $\alpha_t$ [010] | $\alpha_t$ [110] | $\Delta\alpha_t$ |
| 1 | 2.36 | 1.39 | 69.8% |
| 2 | 2.27 | 1.32 | 72.0% |
| 3 | 2.22 | 1.26 | 76.2% |
| 4 | 2.25 | 1.26 | 78.6% |
| 5 | 2.42 | 1.38 | 75.4% |
| **Average** | 2.30 | 1.32 | 74.2% |

Table S5. Onsite damping ($\alpha_{onsite} \times 10^{-3}$) of a typical atom in each layer of the VCA Fe$_{50}$Co$_{50}$(001) bcc for the spin quantization axis [010] and [110].

| Layer | $\alpha_{onsite}$ [010] | $\alpha_{onsite}$ [110] | $\Delta\alpha_{onsite}$ |
|---|---|---|---|
| 1 | 7.36 | 8.70 | +18.2% |
| 2 | 0.63 | 0.69 | +9.5% |
| 3 | 1.41 | 1.44 | +2.1% |
| 4 | 0.87 | 0.86 | −1.1% |
| 5 | 0.99 | 0.97 | −2.0% |

Table S6. Summary of the main $Fe_{50}Co_{50}$ damping anisotropy results for: pure ordered ($B2$) alloy; pure random (VCA) bulk alloy; bcc bulk together with short-range order (SRO) clusters (see Table S3); bulk together with bct-like distorted clusters inside (see Table S4); surface calculations, in the pristine mode and with explicit bct-like clusters embedded (surface + distortion). The maximum-minimum ratio according to Ref. [22] is $\frac{\alpha_t^{[110]}}{\alpha_t^{[010]}} \times 100\%$.

| Structure | $\Delta\alpha_t$ | Max-min ratio |
|---|---|---|
| Ordered alloy bcc | 7.2% | 107.2% |
| Ordered alloy bct | 24.1% | 124.1% |
| Random alloy bcc | 13% | 113% |
| Random alloy bct | 22.6% | 122.6% |
| Random alloy + SRO | 14.9% | 114.9% |
| Random alloy + SRO + Distortion | 47.2% | 147.2% |
| Surface (external layer) | 102.4% | 202.4% |
| Surface (ext. layer) + Distortion | 75.4% | 175.4% |
| 10-nm $Co_{50}Fe_{50}$/Pt [22] (exp.) | 281.3% | 381.3% |

For the outermost layer of Fe(001) bcc, the calculated LDOS at $E_F$ is $\sim 20.42$ states/Ry-atom in the [110] direction and $\sim 20.48$ states/Ry-atom in the [010] direction, which represents a difference of $\sim 0.3\%$ and agrees with the calculations performed by Chen *et al.* [38].

### V. Correlation with anisotropic orbital moment

Besides the close relation exhibited between $\Delta\alpha_t$ and $\Delta n(E_F)$, we also demonstrate the existence of an anisotropic orbital moment in the outermost layer, in which the fourfold symmetry ($C_{4v}$) matches the damping anisotropy with a 90° phase. Fig. S3 shows this correlation between $\Delta\alpha_t$ and $\Delta m_{orb}$ for two situations: ($i$) for a typical atom in the outermost layer of VCA $Fe_{50}Co_{50}$(001) bcc (blue open dots); and ($ii$) for a typical atom in the VCA $Fe_{50}Co_{50}$ bcc bulk, considering the same $\Delta m_{orb}$ scale. For case ($i$) we find orbital moments differences more than one order of magnitude higher than case ($ii$).

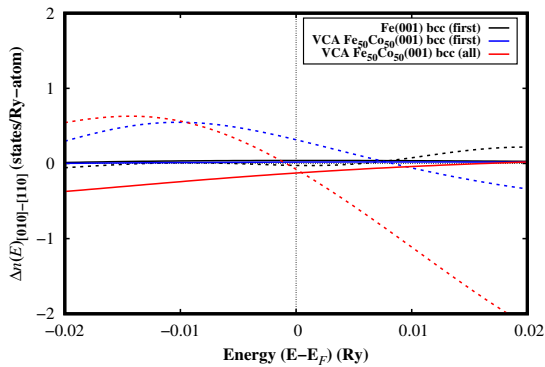

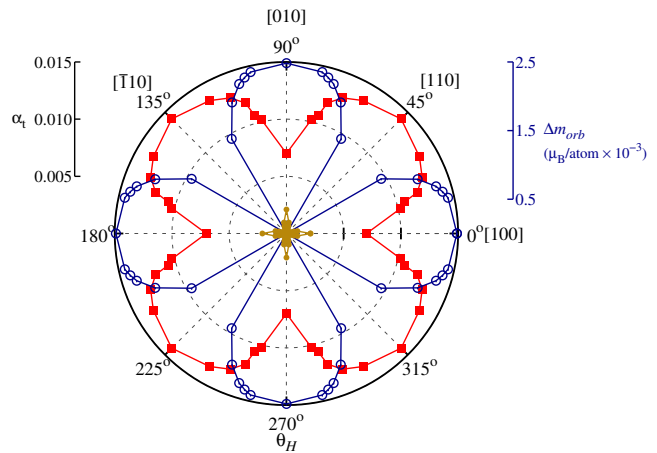

Figure S2. LDOS difference (per atom), $\Delta n(E)$, between the [010] and [110] magnetization directions, for both spin channels (full lines for majority spin and dashed lines for minority spin states), in the outermost layer in pure Fe(001) bcc (in black); outermost layer in VCA $Fe_{50}Co_{50}$(001) bcc (in blue); and all layers summed in VCA $Fe_{50}Co_{50}$(001) bcc (in red).

level, $N(E_F)$, what the authors claim that could not explain the giant maximum-minimum damping ratio observed. So, in order to clarify this effect in the VCA $Fe_{50}Co_{50}$(001) bcc, $\Delta n(E)$ was also calculated for the all layers summed, what is shown in Fig. S2 (in red). This difference is in fact smaller if we consider the DOS of the whole system, with all layers summed. However, if we consider only the most external layer, then the LDOS variation is enhanced. This is consistent with our theoretical conclusions. As we mention in the main text, this do not rule out a role also played by local (tetragonal-like) distortions and other bulk-like factors in the damping anisotropy.

Figure S3. (Color online) Total damping and orbital moment difference, $\Delta m_{orb}$ as a function of $\theta_H$, the angle between the magnetization direction and the [100]-axis. Squares: (red full) VCA $Fe_{50}Co_{50}$(001) bcc. Circles: (blue open) $m_{orb}$ difference between $\theta_H = 90°$ and the current angle for a typical atom in the outermost layer of VCA $Fe_{50}Co_{50}$(001) bcc; and (yellow full) same $m_{orb}$ difference but for a typical atom in the VCA $Fe_{50}Co_{50}$ bcc bulk (in the same scale). Lines are guides for the eyes.

### VI. Contribution from next-nearest-neighbours

Finally, we show in Fig. S4 the summation of all non-local damping contributions, $\alpha_{ij}$, for a given normalized distance in the outermost layer of VCA $Fe_{50}Co_{50}$(001) bcc. As we can see, the next-nearest-neighbours from a reference site (normalized distance $\frac{d}{a}=1$) have very distinct $\alpha_{ij}$ contributions to $\alpha_t$ for the two different magnetization directions ([010] and [110]), playing an important role on the final damping anisotropy. We must note, however, that these neighbours in a (001)-oriented bcc surface are localized in the same layer as the reference site, most affected by the interfacial SOC. Same trend is observed for $\frac{d}{a}=2$, however less intense. This is consistent with our conclusions, about the relevance of the outermost layer on $\Delta\alpha_t$.

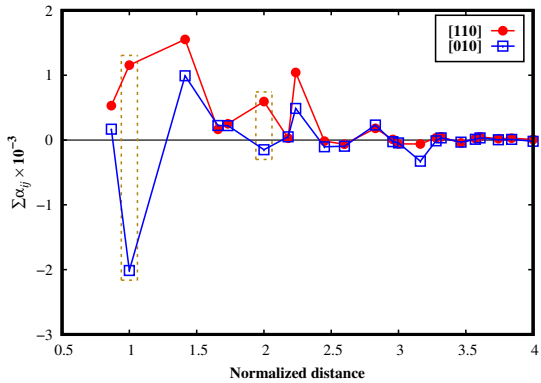

Figure S4. (Color online) Summation of all non-local Gilbert damping parameters ($\alpha_{ij}$, $i \neq j$) in each neighboring normalized distance between sites $i$ and $j$ for the VCA $Fe_{50}Co_{50}$(001) bcc in the two most different directions for the damping anisotropy: [010] ($\theta_H = 90°$), in blue open squares, and [110] ($\theta_H = 45°$), in red full circles. Lines are guides for the eyes.